\documentclass[10pt,twocolumn]{article}  

\usepackage{geometry}
\usepackage{titling}  
\pretitle{\begin{center}\bf\LARGE}
\posttitle{\par\end{center}\vspace{-0.25in}}
\postdate{\par\end{center}\vspace{-0.50in}}

\usepackage{afterpage}

\usepackage[12hr]{datetime}

\usepackage{fancyhdr}

\fancypagestyle{plain}{\lhead{}\chead{}\rhead{}\rfoot{}

}
\usepackage{etoolbox}

\usepackage[multiple]{footmisc}

\usepackage{abstract}
\usepackage[T1]{./fontenc} 

\usepackage[scaled]{helvet}    
\newcommand{\hel}{\fontfamily{phv}\selectfont} 									

\usepackage{setspace}     
\usepackage{indentfirst}  
\usepackage{xspace}       
\usepackage{rotating}     
\usepackage[switch]{lineno}   

\usepackage{graphicx}   
\usepackage{wrapfig}    

\usepackage{booktabs}   
\usepackage{multirow}   
\usepackage{arydshln}   
\usepackage[table]{xcolor}
\usepackage{tabularx}   %
\newcolumntype{R}{>{\raggedleft\arraybackslash}X}

\usepackage{listings}
\usepackage{algorithm}  
\usepackage[noend]{algpseudocode}
\usepackage{algorithmicx}
\floatname{algorithm}{Template}

\usepackage[labelsep=period, labelfont=bf]{caption}
\usepackage[labelsep=period, labelformat=simple, font={normal}]{subcaption}

\usepackage{enumitem}   
\setlist{partopsep=0pt, topsep=4pt, itemsep=4pt, parsep=0pt}

\usepackage{amsmath}    

\usepackage[text-rm]{siunitx}
\mathchardef\mhyphen="2D

\usepackage{amsthm}

\usepackage{verbatim}     
\newcommand{\ignore}[1]{} 

\usepackage{bibspacing}
\usepackage{url}  

\usepackage[compact]{titlesec}

\titlelabel{\thetitle.\hspace{0.3em}}
\titleformat*{\section}{\bf\fontsize{12pt}{12pt}\selectfont}
\titleformat*{\subsection}{\bf\fontsize{11pt}{11pt}\selectfont}
\titleformat*{\subsubsection}{\fontsize{10pt}{10pt}\selectfont}

\usepackage{stmaryrd}

\usepackage{bbding}

\usepackage{pifont} 

\title{\vspace{3mm} RowHammer: Reliability Analysis and Security Implications \vspace{3mm}}

\date{}

\usepackage{authblk}
\usepackage{hyperref}

\makeatletter
\renewcommand\AB@affilsepx{\qquad\protect\Affilfont}
\makeatother

\author[1]{Yoongu Kim}
\author[ ]{Ross Daly}
\author[1]{Jeremie Kim}
\author[ ]{Chris Fallin}
\author[1]{Ji Hye Lee}
\author[1]{\authorcr Donghyuk Lee}
\author[2]{Chris Wilkerson}
\author[ ]{Konrad Lai}
\author[1]{Onur Mutlu}
\affil[1]{{\em Carnegie Mellon University}}
\affil[2]{{\em Intel Labs}}

\newcommand{\reqx}{$\mathit{req}_\mathtt{X}$}
\newcommand{\reqy}{$\mathit{req}_\mathtt{Y}$}
\newcommand{\actx}{$\mathtt{ACT}_\mathtt{X}$}
\newcommand{\acty}{$\mathtt{ACT}_\mathtt{Y}$}
\newcommand{\rdx}{$\mathtt{RD}_\mathtt{X}$}
\newcommand{\rdy}{$\mathtt{RD}_\mathtt{Y}$}
\newcommand{\prex}{$\mathtt{PRE}_\mathtt{X}$}
\newcommand{\prey}{$\mathtt{PRE}_\mathtt{Y}$}

\begin{document}
\maketitle

\fancypagestyle{yoongustyle}{
\chead{This is the summary of the paper titled ``Flipping Bits in Memory
Without Accessing Them: An Experimental Study of DRAM Disturbance
Errors'' which appeared in ISCA in June 2014~\cite{kim14}.}
}
\thispagestyle{yoongustyle}


\setstretch{0.91}  


\section{RowHammer: A New DRAM Failure Mode}

As process technology scales down to smaller dimensions, DRAM chips
become more vulnerable to {\em disturbance}, a phenomenon in which
different DRAM cells interfere with each other's operation. For the
first time in academic literature, our ISCA paper~\cite{kim14} exposes
the existence of {\em disturbance errors} in commodity DRAM chips that
are sold and used today. We show that repeatedly reading from the same
address could corrupt data in nearby addresses. More specifically:

\vspace{-3pt}

\begin{quote} {\em When a DRAM row is opened (i.e., activated) and
closed (i.e., precharged) repeatedly (i.e., {\em hammered}), it can induce
disturbance errors in adjacent DRAM rows.} \end{quote}

\vspace{-3pt}

This failure mode is popularly called {\em RowHammer}. We tested 129
DRAM modules manufactured within the past six years (2008--2014) and
found 110 of them to exhibit RowHammer disturbance errors, the
earliest of which dates back to 2010. In particular, {\em all} modules
from the past two years (2012--2013) were vulnerable, which implies
that the errors are a recent phenomenon affecting more advanced
generations of process technology. Importantly, disturbance errors
pose an easily-exploitable {\em security threat} since they are a
breach of memory protection, wherein accesses to one page (mapped to
one row) modifies the data stored in another page (mapped to an
adjacent row).

Our ISCA paper~\cite{kim14} makes the following contributions.

\begin{enumerate}[leftmargin=1.2em]

    \item We {\bf demonstrate} the existence of DRAM disturbance
      errors on real DRAM devices from three major manufacturers and
      real systems using such devices (using a simple piece of
      user-level assembly code).

    \item We {\bf characterize} in detail the characteristics and
      symptoms of DRAM disturbance errors using an FPGA-based DRAM
      testing platform.

    \item We propose and explore various soluions to {\bf prevent}
      DRAM disturbance errors. We develop a novel, low-cost
      system-level approach as a viable solution to the RowHammer
      problem.

\end{enumerate}

\subsection{Demonstration of the RowHammer Problem}



Code~1a is a short piece of assembly code that we constructed to induce
DRAM disturbance errors on real systems. It is designed to generate a
read to DRAM on every data access. First, the two {\tt mov} instructions
read from DRAM at address {\tt X} and {\tt Y} and install the data into
a register and also the cache. Second, the two {\tt clflush}
instructions evict the data that was just installed into the cache.
Third, the {\tt mfence} instruction ensures that the data is fully
flushed before any subsequent memory instruction is executed. Finally,
the code jumps back to the first instruction for another iteration of
reading from DRAM.

On processors employing out-of-order execution, Code~1a generates
multiple read requests, all of which queue up in the memory controller
before they are sent out to DRAM: \reqx, \reqy, \reqx, \reqy,
$\cdots$. Importantly, we chose the values of {\tt X} and {\tt Y} so
that they map to {\em different} rows within the {\em same} bank. This
is so that the memory controller is forced to open and close the two
rows repeatedly: \actx, \rdx, \prex, \acty, \rdy, \prey,
$\cdots$. Using the address-pair ({\tt X}, {\tt Y}), we then executed
Code~1a for millions of iterations. Subsequently, we repeated this
procedure using many different address-pairs until every row in the
DRAM module was opened/closed millions of times. In the end, we
observed that Code~1a caused many bits to flip. For four different
processors, Table~\ref{tab:memtest} reports the total number of
bit-flips induced by Code~1a for two different initial states of the
module: all `0's or all `1's. Since Code~1a {\em does not write} any
data into DRAM, we conclude that the bit-flips are the manifestation
of disturbance errors caused by repeated reading (i.e., hammering) of
each memory row. In the next section, we will show that this
particular DRAM module yields {\em millions} of errors in a more
controlled environment.

\vspace{1ex}
{
\noindent
\textcolor{gray}{\rule[-21pt]{2pt}{67pt}}\begin{minipage}{0.47\linewidth}
\floatname{algorithm}{}
\captionsetup[algorithm]{labelsep=none}
\renewcommand{\thealgorithm}{}
\label{algo:exploit1}
\begin{algorithmic}[1]
\algrenewcommand\alglinenumber[1]{\footnotesize\texttt{#1}}
\small
\tt
\State \underline{code1a}:
\State \ \ mov (X), \%eax
\State \ \ mov (Y), \%ebx
\State \ \ clflush (X)
\State \ \ clflush (Y)
\State \ \ mfence
\State \ \ jmp \underline{code1a}
\end{algorithmic}
\captionof{algorithm}{\hspace{-0em}{\bf a.}~Induces errors}

\end{minipage}\hspace{6pt}%
\textcolor{gray}{\rule[-21pt]{2pt}{67pt}}\begin{minipage}{0.47\linewidth}
\floatname{algorithm}{}
\captionsetup[algorithm]{labelsep=none}
\renewcommand{\thealgorithm}{}
\label{algo:exploit2}
\begin{algorithmic}[1]
\algrenewcommand\alglinenumber[1]{\footnotesize\texttt{#1}}
\small
\tt
\State \underline{code1b}:
\State \ \ mov (X), \%eax
\State \ \ clflush (X)
\State \ \ 
\State \ \ 
\State \ \ mfence
\State \ \ jmp \underline{code1b}
\end{algorithmic}
\captionof{algorithm}{\hspace{-0em}{\bf b.}~Does not induce errors}

\end{minipage}
\floatname{algorithm}{Code}%
\renewcommand{\thealgorithm}{1}%
\captionof{algorithm}{Assembly code executed on Intel/AMD machines}
\vspace{6pt}
}

\begin{table}[h]
\centering
\small

\setlength{\tabcolsep}{5pt}
\sisetup{group-separator={,}}
\sisetup{group-minimum-digits=4}
\begin{tabular}{c*{4}{S[table-format=5]}}

\toprule


\multirow{2}{*}{Bit-Flip} & \text{Intel} & \text{Intel} & \text{Intel} & \text{AMD} \\
                          & \text{Sandy Bridge} & \text{Ivy Bridge} & \text{Haswell} & \text{Piledriver} \\

\midrule

`0' $\shortrightarrow$ `1'     & 7992 & 10273 & 11404 & 47 \\


`1' $\shortrightarrow$ `0'     & 8125 & 10449 & 11467 & 12 \\



\bottomrule

\end{tabular}
\caption{Bit-flips induced by disturbance on a 2GB module} 
\label{tab:memtest}
\end{table}

As a control experiment, we also ran Code~1b which reads from only a
single address. Code~1b did {\em not} induce any disturbance errors as
we expected. For Code~1b, all of its reads are to the same row in
DRAM: \reqx, \reqx, \reqx, $\cdots$. In this case, the memory
controller minimizes the number of DRAM commands~\cite{rixner00,
  zuravleff97, stfm, parbs, tcm, atlas, sms, bliss, mise} by opening
and closing the row just {\em once}, while issuing many column reads
in between: \actx, \rdx, \rdx, \rdx, $\cdots$, \prex. From this we
conclude that DRAM disturbance errors are indeed caused by the
repeated opening/closing of a row, and {\em not} by the reads
themselves.


\subsection{Characterization of the RowHammer Problem}

To develop an understanding of disturbance errors, we characterize 129
DRAM modules on an FPGA-based DRAM testing
platform~\cite{retentionfail, aldram, rowhammer, kim14,
  avatar}. Unlike a general-purpose processor, our testing platform
grants us precise and fine-grained control over how and when DRAM is
accessed on a cycle-by-cycle basis. To characterize a module, we test
each one of its rows one by one. First, we initialize the entire
module with a known data-pattern. Second, we activate one particular
row as quickly as possible (once every 55{\em ns}) for the full
duration of a refresh interval (64{\em ms}). Third, we read out the
entire module and search for any changes to the data-pattern. We then
repeat the three steps for every row in the module.

{\bf Key Findings.} In the following, we summarize four of the most
important findings of our RowHammer characterization study.

\begin{figure}[h]
\centering
\includegraphics{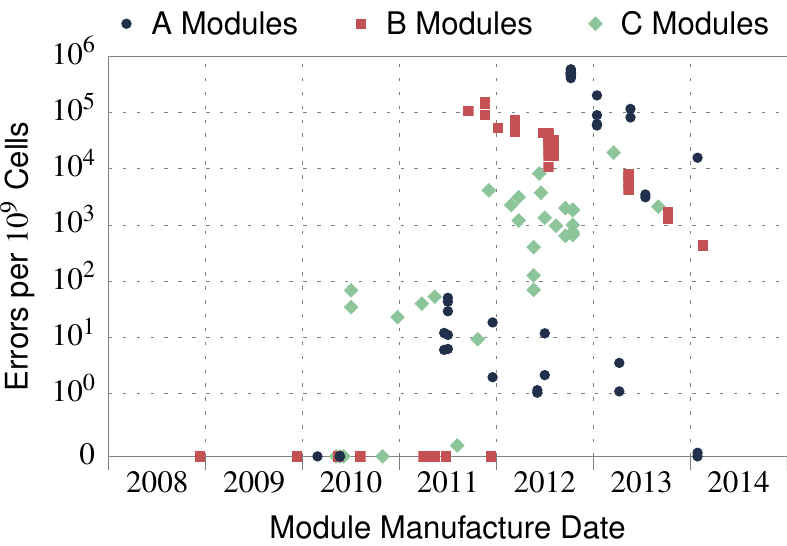}
\caption{Normalized number of errors vs.~manufacture date}
\label{fig:errors_vs_date}
\end{figure}

{\em 1.~Errors are widespread.} Figure~\ref{fig:errors_vs_date} plots
the normalized number of errors for each of the 129 modules as a
function of their manufacture date. Our modules are sourced from three
major DRAM manufacturers whose identities have been anonymized to {\hel
A}, {\hel B}, and {\hel C}. From the figure, we see that disturbance
errors first started to appear in 2010, and that they afflict all
modules from 2012 and 2013. In particular, for each manufacturer, the
number of errors per $10^9$ cells can reach up to $5.9 \times 10^5$,
$1.5 \times 10^5$, and $1.9 \times 10^4$, respectively. To put this into
perspective, there can be as many as 10 million errors in a 2GB module.

{\em 2.~Errors are symptoms of charge loss.} For a given DRAM cell, we
observed that it experiences data loss in only a single direction:
either `1'$\shortrightarrow$`0' or `0'$\shortrightarrow$`1', but not
both. This is due to an intrinsic property of DRAM cells called {\em
orientation}. Depending on the implementation, some cells represent a
data value of `1' using the charged state, while other cells do so using
the discharged state --- these cells are referred to as {\em true-cells}
and {\em anti-cells}, respectively~\cite{retention}. We profiled several
modules for the orientation of their cells, and discovered that
true-cells experience only `1'$\shortrightarrow$`0' errors and that
anti-cells experience only `0'$\shortrightarrow$`1' errors. From this,
we conclude that disturbance errors occur as a result of charge loss.

{\em 3.~Errors occur in adjacent rows.} We verified that the disturbance
errors caused by activating a row are localized to two of its
immediately adjacent rows. There could be three possible ways in which a
row interacts with its neighbors to induce their charge loss: {\em (i)}
electromagnetic coupling, {\em (ii)} conductive bridges, and {\em (iii)}
{\em hot-carrier injection}. We confirmed with at least one major DRAM
manufacturer that these three phenomena are potential causes of the
errors. Section 3 of our ISCA paper~\cite{kim14} provides more analysis.

{\em 4.~Errors are access-pattern dependent.} For a cell to experience a
disturbance error, it must lose enough charge before the next time it is
replenished with charge (i.e., {\em refreshed}). Hence, the more
frequently we refresh a module, the more we counteract the effects of
disturbance, which decreases the number of errors. On the other hand,
the more frequently we activate a row, the more we strengthen the
effects of disturbance, which increases the number of errors. We
experimentally validated this trend by sweeping the {\em refresh
interval} and the {\em activation interval} between 10--128{\em ms} and
55--500{\em ns}, respectively. In particular, we observed that no errors
are induced if the refresh interval is $\le$8{\em ms} or if the
activation interval is $\ge$500{\em ns}. Importantly, we found that it
takes as few as 139K activations to a row before it induces disturbance
errors.

{\bf Other Findings.} Our ISCA paper provides an extensive set of
characterization results, some of which we list below.

\begin{itemize}[leftmargin=1.2em]

    \item RowHammer errors are repeatable. Across ten iterations of
      tests, >70\% of the erroneous cells had errors in every
      iteration.

    \item Errors are not strongly affected by temperature. The numbers
        of errors at 30$^\circ$C, 50$^\circ$C, and 70$^\circ$C differ by
        <15\%.

    \item There is almost no overlap between erroneous cells and {\em
      weak cells} (i.e., cells that are inherently the leakiest and
      thus require a higher refresh rate).

    \item Simple ECC (e.g., SECDED) {\em cannot} prevent {\em all}
      RowHammer-induced errors.  There are as many as four errors in a
      single cache-line.

    \item Errors are data-pattern dependent. The {\em Solid}
        data-pattern (all `0's or all `1's) yields the fewest errors,
        whereas the {\em RowStripe} data-pattern (alternating rows of
        `0's and `1's) yields the most errors.

    \item A very small fraction of cells experience an error when either
        one of their adjacent rows is repeatedly activated.

\end{itemize}

\subsection{Prevention of the RowHammer Problem}

In our ISCA paper, we examine a total of {\em seven} solutions to
tolerate, prevent, or mitigate DRAM disturbance errors. The first six
solutions are: 1) making better DRAM chips, 2) using error correcting
codes (ECC), 3) increasing the refresh rate, 4) remapping error-prone
cells after manufacturing, 5) remapping/retiring error-prone cells at
the user level during operation, 6) identifying hammered rows and
refreshing their neighbors. None of these frst six solutions are very
desirable as they come at various significant power, performance or
cost overheads.

Our main proposal to solve the RowHammer problem is a novel
low-overhead mechanism called {\em PARA} ({\em probabilistic adjacent
  row activation}). The key idea of PARA is simple: every time a row
is opened and closed, one of its adjacent rows is also opened (i.e.,
refreshed) with some low probability. If one particular row happens to
be opened and closed repeatedly, then it is statistically certain that
the row's adjacent rows will eventually be opened as well.  The main
advantage of PARA is that it is {\em stateless}. PARA does not require
expensive hardware data-structures to count the number of times that
rows have been opened or to store the addresses of the error-prone
cells. PARA can be implemented either in the memory controller or the
DRAM chip (internally).

PARA is implemented as follows. Whenever a row is closed, the PARA
control logic flips a biased coin with a probability $p$ of turning up
heads, where $p \ll 1$. If the coin turns up heads, the controller
opens one of the adjacent rows where either of them is chosen with
equal probability ($p/2$). The parameter $p$ can be set so that
disturbance errors occur at an extremely low rate --- many orders of
magnitude lower than the failure rates of other system components
(e.g., hard-disk). In fact, even under the most adversarial
conditions, PARA's failure rate is only $9.4 \times 10^{-14}$
errors-per-year when {\em p} is set to just 0.001. Due to the extra
activations, PARA incurs a small performance overhead (slowdown of
0.20\% averaged across 29 benchmarks), which we believe to be
justified by the {\em (i)} strong reliability guarantee and {\em (ii)}
low design complexity resulting from PARA's stateless nature. More
detailed analysis can be found in our ISCA 2014 paper~\cite{kim14}.

\setstretch{0.91}  

\section{Significance}

Our ISCA paper~\cite{kim14} identifies a new reliability problem and a
security vulnerability, {\em RowHammer}, that affects an entire
generation of computing systems being used today. We build a
comprehensive understanding of the problem based on a wealth of
empirical data we obtain from more than 120 DRAM memory modules. After
examining various ways of addressing the problem, we propose a
low-overhead solution that provides a strong reliability (and,
hopefully, security) guarantee.

{\bf Exposition.} For the first time in academic literature, we expose
the widespread vulnerability of commodity DRAM chips to disturbance
(aka, RowHammer) errors. After testing a large sample population of
DRAM modules (the oldest of which dates back to 2008), we determine
that the problem first arose in 2010 and that it still persists to
this day. We found modules from {\em all} three major manufacturers,
as well as in {\em all} modules assembled between 2012--2013, are
vulnerable to the RowHammer problem.

{\bf Demonstration.} We demonstrate that disturbance errors are an
actual hardware vulnerability affecting real systems. We construct a
user-level kernel which induces many errors on general-purpose
processors from Intel (Sandy Bridge, Ivy Bridge, Haswell) and AMD
(Piledriver). With its ability to bypass memory protection (OS/VMM),
the kernel can be deployed as a {\em disturbance attack} to corrupt
the memory state of a system and its software. We discuss the
possibility, for the first time, that this {\em RowHammer} problem can
be developed into a malicious ``disturbance attack'' (Section 4 of our
ISCA paper~\cite{kim14}).\footnote{Recent
  works~\cite{google-project-zero, rowhammer-js} that build upon our
  ISCA 2014 paper actually demonstrate that a user can take over an
  entire system and gain kernel privilges by intelligently exploiting
  the RowHammer problem.}

{\bf Characterization.} We characterize the cause and symptoms of
disturbance errors based on a large-scale study, involving 129 DRAM
modules (972 DRAM chips) sampled from a time span of six years. We
extensively test the modules using a custom-built FPGA
infrastructure~\cite{retentionfail, aldram, retention, kim14, avatar}
to determine the specific conditions under which the errors occur, as
well as the specific manner in which they occur. From this, we build a
comprehensive understanding of disturbance errors.

{\bf Solution.} We propose seven categories of solutions that could
potentially be employed to prevent disturbance errors. Among them, our
main proposal is called {\em PARA} ({\em probabilistic adjacent row
activation}), whose major advantage lies in its stateless nature. PARA
eliminates the need for hardware counters to track the number of
activations to different rows (proposed in other potential
solutions~\cite{bains14c, bains14a, bains14b, bains14d, greenfield14b,
greenfield14a,moinrowhammer}), while still being able to refresh the
at-risk rows in a timely manner. Based on our analysis, we establish
that PARA provides a strong reliability guarantee even under the
worst-case conditions, proving it to be an effective and efficient
solution against the RowHammer problem.

\subsection{Long-Term Impact}


We believe our ISCA paper will affect industrial and academic research
and development for the following four reasons. First, it exposes a
real and pressing manifestation of the difficulties in DRAM scaling
--- a critical problem which is expected to become only worse in the
future~\cite{mutlu13,superfri,kiise}. Second, it breaks the
conventional wisdom of memory protection, demonstrating that the
system-software (OS/VMM) --- just by itself --- cannot isolate the
address space of one process (or virtual machine) from that of
another, thereby exposing the vulnerability of systems, for the first
time, to what we call {\em DRAM disturbance attacks}, or {\em
  RowHammer attacks}.  Third, it builds the necessary experimental
infrastructure to seize full control over DRAM chips, thereby
unlocking a whole new class of characterization studies and
quantitative data. Fourth, it emphasizes the important role of
computer architects to examine holistic approaches for improving
system-level reliability and security mechanisms for modern memory
devices --- even when the underlying hardware is unreliable.

{\bf Empirical Evidence of Challenges in DRAM Technology Scaling.} DRAM process
scaling is becoming more difficult due to increased cost and complexity, as
well as degraded reliability~\cite{cha11, hong10, kang14, mutlu13, superfri,
kiise}. This explains why disturbance errors are found in all three DRAM
manufacturers, in addition to why their first appearances coincide with each
other during the same timeframe (2010--2011). As process scaling continues
onward, we could be faced with a new and diverse array of DRAM failures, some
of which may be diagnosed only after they have been released into the wild ---
as was the case with disturbance errors of today. In this context, our paper
raises awareness about the next generation of DRAM failures that could
undermine system integrity. We expect, based on our experimental evidence of
DRAM errors, that future DRAM chips could suffer from similar or other
vulnerabilities.

{\bf RowHammer Security Implications: Threats of Unreliable Memory.}
Virtual machines and virtual memory protection mechanisms exist to
provide an isolated execution environment that is safeguarded from
external tampering or snooping. However, in the presence of
disturbance errors (or other hardware faults), it is possible for one
virtual machine (or application) to corrupt the memory of another
virtual machine (or application) that is housed in the same physical
machine. This has serious implications for modern systems (from mobile
systems to data centers), where multiprogramming is common and many
virtual machines (or applications) potentially from different users
are consolidated onto the same physical machine. As long as there is
sharing between any two pieces of software, our paper shows that
strong isolation guarantees between them {\em cannot} be provided
unless all levels of the system stack are secured. As a result,
malicious software can be written to take advantage of these
disturbance errors. We call these {\em disturbance attacks}, or {\em
  RowHammer attacks}. Such attacks can be used to corrupt system
memory, crash a system, or take over the entire system. Confirming the
predictions of our ISCA paper~\cite{kim14}, researchers from Google
Project Zero recently developed a user-level attack that exploits
disturbance errors to take over an entire
system~\cite{google-project-zero}. More recently, researchers have
shown that the RowHammer can be exploited remotely via the use of
JavaScript~\cite{rowhammer-js}. As such, the new problem exposed by
our ISCA 2014 paper, the DRAM RowHammer problem, has widespread and
profound real implications on system security, threatening the
foundations of memory security on top of which modern systems are
built.

{\bf DRAM Testing Infrastructure.} Our paper builds a powerful
infrastructure for testing DRAM chips. It was designed to grant the
user with software-based control over the precise timing and the exact
data with which DRAM is accessed. This creates new opportunities for
DRAM research in ways that were not possible before. For example, we
have already leveraged the infrastructure for {\em (i)} characterizing
different modes of retention failures~\cite{khan14, retention} and
{\em (ii)} characterizing the safety margin in timing parameters to
operate DRAM at lower latencies than what is
recommended~\cite{aldram}. In the future, we plan to open-source the
infrastructure for the benefit of other researchers.

{\bf System-Level Approach to Enable DRAM Scaling.} Unlike most other
known DRAM failures, which are relatively easily caught by the
manufacturers, disturbance errors require an extremely large number of
accesses before they are sensitized --- a full-coverage test to reveal
all disturbance errors could take days or weeks. As DRAM cells become
even smaller and less reliable, it is likely for them to become even
more vulnerable to complicated and different modes of failure which
are sensitized only under specific access-patterns and/or
data-patterns. As a scalable solution for the future, our paper argues
for adopting a system-level approach~\cite{mutlu13} to DRAM
reliability and security, in which the DRAM chips, the memory
controller, and the operating system collaborate together to
diagnose/treat emerging DRAM failure modes.

\section{Conclusion}

Our ISCA 2014 paper~\cite{kim14} is the first work that exposes,
demonstrates, characterizes, and prevents a new type of DRAM failure,
the DRAM RowHammer problem, which can cause serious security and
reliability problems. As DRAM process technology scales down to even
smaller feature sizes, we hope that our findings will inspire the
research community to develop innovative approaches to enhance
system-level reliability and security by focusing on such new forms of
memory errors and their implications for modern computing systems.

\section{More Information}

For more information, we point the reader to the following resources:

1. We have released source code to induce DRAM RowHammer errors as
open source software~\cite{rowhammer-source}.

2. We have released presentations on the RowHammer
problem~\cite{rh-talk1,rh-talk2}.

3. We have written various papers describing challenges in DRAM
scaling and memory systems in general~\cite{mutlu13,superfri,kiise}.

4. Building upon our observations, others have exploited the RowHammer
problem to take over modern systems and have released their source
code~\cite{google-project-zero,rowhammer-js}.

5. Mark Seaborn maintains a discussion group~\cite{rh-discuss} that
discusses RowHammer issues.

6. More detailed background information and discussion on DRAM can be
found in our video lectures~\cite{dram-lecture1,dram-lecture2} or
recent works that explains the operation and architecture of modern
DRAM
chips~\cite{raidr,salp,tldram,aldram,retentionfail,retention,darp,rowclone,ramulator}.

7. Twitter has a record of popular discussions on the RowHammer
problem~\cite{rh-twitter}.


\setstretch{0.86}

\titleformat*{\section}{\bf\fontsize{12pt}{12pt}\selectfont}
\vspace{10pt}
{\footnotesize
\bibliography{paper}
}


\end{document}